\documentclass[preprint,12pt]{elsarticle}

\usepackage{graphicx}
\usepackage{longtable}

\usepackage{amssymb}

\biboptions{numbers,sort&compress}

\journal{Physica A}

\begin{document}

\begin{frontmatter}



\title{Allometric Scaling of Countries}


\author{Jiang Zhang, Tongkui Yu}

\address{Department of Systems Science, School of Management, Beijing Normal University, Beijing, 100875}
\ead{zhangjiang@bnu.edu.cn}

\begin{abstract}
As huge complex systems consisting of geographic regions, natural
resources, people and economic entities, countries follow the
allometric scaling law which is ubiquitous in ecological, urban
systems. We systematically investigated the allometric scaling
relationships between a large number of macroscopic properties and
geographic (area), demographic (population) and economic (GDP, gross
domestic production) sizes of countries respectively. We found that
most of the economic, trade, energy consumption, communication
related properties have significant super-linear (the exponent is
larger than 1) or nearly linear allometric scaling relations with
GDP. Meanwhile, the geographic (arable area, natural resources,
etc.), demographic(labor force, military age population, etc.) and
transportation-related properties (road length, airports) have
significant and sub-linear (the exponent is smaller than 1)
allometric scaling relations with area. Several differences of power
law relations with respect to population between countries and
cities were pointed out. Firstly, population increases sub-linearly
with area in countries. Secondly, GDP increases linearly in
countries but not super-linearly as in cities. Finally, electricity
or oil consumptions per capita increases with population faster than
cities.
\end{abstract}


\end{frontmatter}


\section{Introduction}
\label{sec.introduction}

Scientists always look for universal patterns of complex systems. As
huge complex systems consisting of geographic regions, natural
resources, people and economic entities, countries also share some
common features despite their different latitudes, climates and
cultures. For example, bigger countries in area always have greater
population; richer countries always provide more healthy lives for
people\citep{Weil2005}. Nevertheless, these features are just
qualitative, vague and subjective judgements. The quantitative
universal patterns which are reflected by the mathematical relations
between macro-variables such as GDP, population, total oil
consumptions etc. are very necessary for macroeconomics of
countries\citep{Weil2005,Blanchard2000}.

Scaling property is one of the most important universal quantitative
laws governing different dynamics of complex
systems\citep{Barenblatt2003,Brock1999,Brown2000}. Allometric
scaling particularly describes the power law relation between two
variables\citep{naroll1956}. In biology, the allometric scaling
relation between metabolism and body size is also known as Kleiber's
law\citep{Brown2000,kleiber1932,west2005}. Biologists found not only
metabolism but also other important biological variables including
heart beat frequency, life span, fertility rate and so on all
exhibit power law relationships with the body size of
species\citep{Brown2000, Brown2004}.

Allometric scaling law can be also extended to the larger complex
systems such as cities \citep{Isalgue2007,
Bettencourt2007,Bettencourtwest2007,Kuhnert2006,Lee1989}. The power
law relation between area and population is one of the earliest
allometries found in cities \citep{Lee1989}. Recently, more scaling
relations were discovered due to more data of cities are available.
The scaling relation between energy consumption and population which
is a parallel of Kleiber's law is discovered for cities
\citep{Isalgue2007, Kuhnert2006, Lammer2006}. Not only energy
consumption but also GDP has a power law relation with population of
cities with an exponent around $1.20$. In
\cite{Bettencourtwest2007}, authors systematically studied
allometric scaling relations between various properties including
inventors, total wages, GDP, total housing, etc. and population of
cities. All of these properties fall into three groups according to
the exponents. The economic and innovation related variables have
super-linear relationship with population (i.e., the exponent is
larger than 1); the household properties have nearly linear
relationship (i.e., the exponent is approaching 1), and the energy
consumption and infrastructure-related properties have sub-linear
relationship (i.e., the exponent is smaller than 1). Allometries
between population and GDP for cities in a province have been also
studied\citep{chen2003}. The exponent in this special allometric
scaling relation is larger than 1.

Although less attention was paid on the allometric scaling of
countries than cities, some existing studies should be noted.
Roehner has studied the power law relationship between the export
value and population of countries\citep{Roehner1984}. He found a
$3/4$ power law relation by grouping countries in different
categories according to GDP per capita. Miguel \citep{Miguel2009}
studied the $CO_2$ emission of countries as the metabolism of the
whole country and compared to the biological allometries.

This paper will try to find the possible universal allometric
scaling of countries by systematic exploring. It is organized as
follows. The first part introduces the methods of our systematic
exploring. The results are discussed in section \ref{sec.results}.
We will show different allometries between variables and area as the
geographic size, population as the demographic size and GDP as the
economic size of countries in section \ref{sec.areasize},
\ref{sec.population} and \ref{sec.GDP} respectivetly. Section
\ref{sec.PopulationGDP} mainly discusses the power law relation
between population and GDP. In section \ref{sec.intensive}, we also
investigate the allometric scaling relationships between intensive
variables. We discussed the major axis regression method and the
comparison between cities and countries in section
\ref{sec.discussion}

\section{Methods}
\label{sec.methods}

We investigated the possible allometric scaling relationships
between macro-variables of countries by empirical data. The main
source of the data is from the software Mathematica. An online
database with hundreds of properties for various countries around
the world is accessible through the embedded command "CountryData"
in Mathematica. These data is from the authoritative organizations,
magazines, year books and websites such as Encyclopaedia Britannica,
Britannica Book of the Year, Encyclopaedia Britannica,United States
National Geospatial-Intelligence Agency, GeoHive: Global Statistics,
World Health Organization, etc.\citep{mathematicacountrydata}. The
carbon emission data is from United Nations Framework Convention on
Climate Change web site \citep{carbonemmission}.

We systematically investigate totally $104$ properties of $273$
countries or regions. Firstly, these properties are classified as
extensive variables ($87$ variables) and intensive variables ($17$
variables). Extensive ones are the gross quantities of countries
such as total area, GDP etc. Intensive variables are the rates or
ratios of several extensive variables such as GDP per capita, birth
rate etc. All variables are classified into $11$ different
categories including geographical properties, natural resources
related properties etc. The variables with their categories,
features,demonstrations are listed in \ref{sec.appendix}.

The allometric scaling law always indicates power law relationship
between macro-variable and the size of the system. For living
organism, the size of the system is its body size or body
mass\citep{Brown2004}. For cities, population is always used as the
measurement of city size \citep{Bettencourtwest2007,Zantte2007}.
However, it is not clear what property is the size of a country. We
can say the size of a country in three different senses, area,
population or GDP. Thus, we studied the allometric scalings of
countries between the given extensive macro-variables and the
geographic size(area), demographic size(population) and economic
size (GDP) of countries separately. We suppose the following
equation holds:
\begin{equation}
\label{eqn.allometricscaling} X=cM^\alpha
\end{equation}

Where $X$ is one of the extensive macro-variables, $M$ is the area
or population or GDP of a country. $c$ and $\alpha$ are numbers we
should estimate. Where, $\alpha$ is the power law exponent of the
allometry which is the most important parameter. We suppose that the
equation holds for all countries. Thus we use linear regression
method to find the best fitting line on log-log coordinates, and
estimate the parameters $c$ and $\alpha$. From equation
\ref{eqn.allometricscaling}, another equation can be derived:
\begin{equation}
\label{eqn.allometrymeaning} X/M=cM^{\alpha-1}
\end{equation}

Note that the left hand of the equation $X/M$ is the average value
of $X$ per size. For example, if $X$ is GDP, $M$ is population, then
$X/M$ is GDP per capita. So if $\alpha>1$, the property $X$ has a
super linear relation with respect to $M$, i.e. GDP per capita
increases with the size $M$. On the other hand, $\alpha<1$ means $X$
has a sub linear relation with respect to $M$, the average value
$X/M$ decreases as the size of country $X$ increases. And the linear
relationship can be got with $\alpha\approxeq
1$\citep{Bettencourtwest2007}.

\begin{figure}
\centerline {\includegraphics{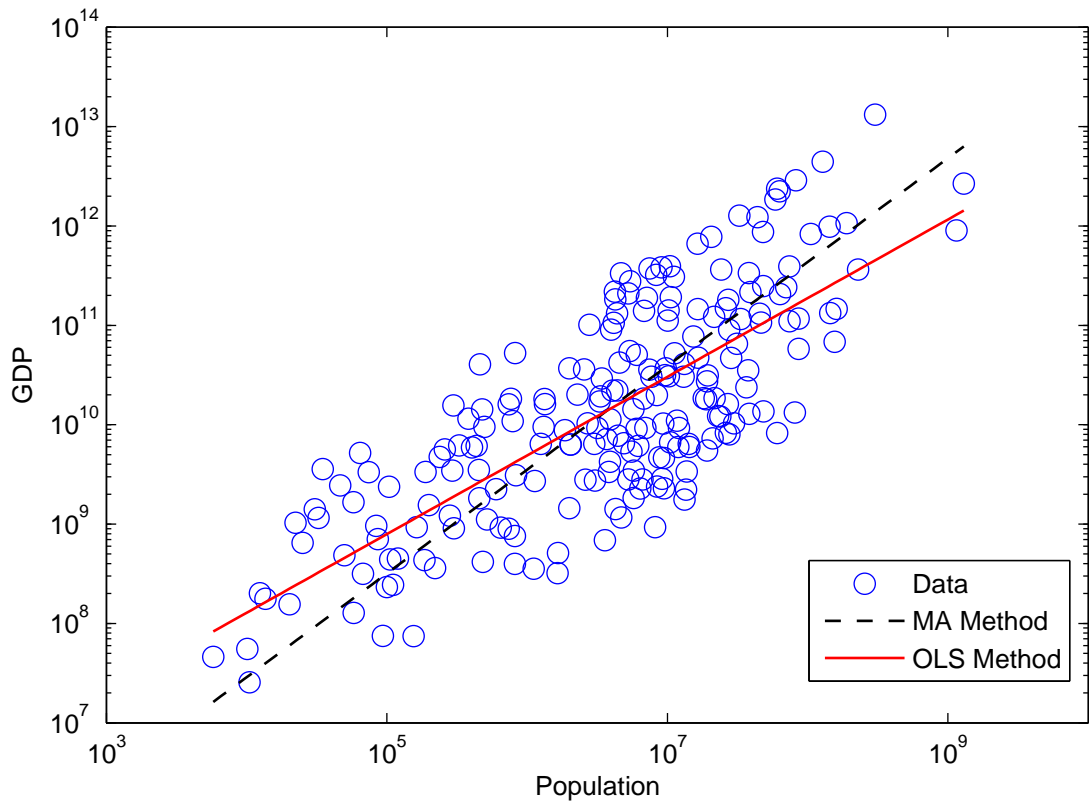}} \vskip3mm \caption{
Different Regression Methods Applied on Population and GDP Power Law
Relation, $\alpha=1.04$ and $\alpha'=0.79$}\label{fig.methods}
\end{figure}

We adopt major axis (MA) approach instead of ordinary least square
(OLS) regression to make linear regression on log-log
coordinate\citep{warton2006,kaitaniemi2004,sokal1995}. MA method is
preferable because our main purpose in this paper is to summarize
the relationship between $M$ and $X$ but not predict $X$ according
to $M$ \citep{warton2006,kaitaniemi2004}. And OLS always give biased
estimations. From figure \ref{fig.methods}, we can see the obvious
difference between slopes estimated by different methods. The line
regressed by MA method locates in the middle of the data band
because this method tries to minimize the total distances from data
points to the regression line both from $x$ and $y$ directions, but
not only the $y$ direction as OLS method does
\citep{warton2006,kaitaniemi2004}. However, we both list exponent
($\alpha$) estimated by MA method and $\alpha'$ by OLS method in
tables \ref{tab.area},\ref{tab.population},\ref{tab.gdp} and
\ref{tab.intensive} to compare these two methods and their predicted
exponents.

We systematically investigate the power law relations between any
given property of the $87$ extensive variables and three kinds of
sizes. The $87\times 3=261$ regressions are computed. The
regressions with significant goodness (the coefficient of
determination $r^2$ is higher than 0.6, where $r$ is calculated as
the correlation coefficient between two variables $X$ and $M$
\citep{sokal1995}) are shown in tables in next section. We also used
the similar method to investigate the regressions of intensive
variables with respect to GDP per capita, the results with $r^2$
larger than $0.4$ are kept. Because some of the properties of some
countries may be absent, the number of observations is smaller than
$273$ in some regressions.

\section{Results}
\label{sec.results}
\subsection{Allometric Scaling with respect to Area}
\label{sec.areasize}

Countries are geometric regions that have relatively clear
boundaries in the geographic spaces, therefore, area is an important
property which can be treated as the geographic size of the
countries. We denote area as $M$, and other properties as $X$ to
regress on log-log coordinate with the methods mentioned in the
previous section. The results are shown in table \ref{tab.area}.

\begin{table}
\centering
 \caption{Allometric Scaling Relations with respect to
 Area(Exponent $\alpha$ is estimated by MA method and $\alpha'$ is
 estimated by OLS method. Only the allometries with $r^2>0.6$ are shown here)}
 \label{tab.area}
\begin{tabular}{ccccccc}
\hline
Property & $\alpha$  & $r^2$ & $\alpha'$ & Observations\\
\hline

BoundaryLength&$0.52\pm0.02$&0.88&0.50&237\\
ArableLandArea&$1.06\pm0.07$&0.81&0.95&218\\
NaturalResources&$0.20\pm0.02$&0.62&0.20&237\\
\\
Population&$0.77\pm0.07$&0.71&0.67&212\\
\\
MilitaryAgePopulation&$0.75\pm0.09$&0.64&0.63&162\\
\\
LaborForce&$0.84\pm0.08$&0.68&0.72&226\\
AgriculturalValueAdded&$0.79\pm0.07$&0.70&0.69&207\\
\\
Airports&$0.57\pm0.04$&0.73&0.52&231\\
RoadLength&$0.78\pm0.05$&0.78&0.71&227\\

\hline
\end{tabular}
\end{table}

Notice that all exponents in table \ref{tab.area} are smaller than 1
except Arable Land Area which should be proportional to areas of
countries. That means all properties have sub-linear allometries
with respect to geometric size. As the area increases, the average
value of population and number of natural sources etc. per area
decreases because more barren places can be observed in countries.
The sub-linear allometries with respect to area implies the
expansion of territory can not give countries proportional returns.

The first property, boundary length of the countries has an about
$1/2$ power law relation with area. This is the obvious geometric
fact which can prove the validity of our method. The allometry
between the number of different natural resources and area can be
compared to the area-species relationship in ecology
\citep{MacArthur1967}. The number of different natural resources in
a country which corresponds to the number of different species found
in an island indicates the diversity of resources. And the area of
country corresponds to the area of island. Therefore, this power law
relationship is the extension of area-species relation in country
scale. Ecologists found that the power law exponent of species-area
relation is from 0.15 to 0.4\citep{Hubbell2001} which is coincident
to the number of natural resource-area power law here.

Another interesting fact is the allometric scaling relation between
population and area of countries. The exponent of this allometry is
approaching $3/4$ which implies a fractal pattern of human being
dwelling. This allometry reminds us the allometric scaling of river
basins\citep{banavar1999,dreyer2001,Rodriguez1997}. It is
interesting to compare the population-area allometry in cities and
countries. Studies on cities showed that the exponent was larger
than 1\citep{Lee1989}, which means the population density
(population per area) increases with city area. Nevertheless, the
allometry with the exponent being smaller than 1 suggests that the
population density decreases with areas of countries. This
conclusion is coincident to our observations that big countries
always have less dense population.

\subsection{Allometric Scaling with respect to Population}
\label{sec.population}

Population is an important property for allometries in cities. We
also studied the allometric scaling relations between population and
other properties of countries. The results are shown in table
\ref{tab.population}.
\begin{center}
\small
\begin{table}
\centering
 \caption{Allometric Scaling Relations with respect to Population(Exponent $\alpha$ is estimated by MA method and $\alpha'$ is
 estimated by OLS method). Only the allometries with $r^2>0.6$ are shown here}
 \label{tab.population}
\begin{tabular}{ccccccc}
\hline
Property & $\alpha$  & $r^2$ & $\alpha'$ & Observations\\
\hline
Area&$1.30\pm0.11$&0.71&1.05&212\\
BoundaryLength&$0.61\pm0.07$&0.61&0.53&212\\
\\
ArableLandArea&$1.31\pm0.08$&0.82&1.16&205\\
IrrigatedLandArea&$1.66\pm0.19$&0.63&1.20&173\\
\\
MilitaryAgePopulation&$1.00\pm0.02$&0.99&1.00&161\\
\\
GDPAtParity&$1.03\pm0.08$&0.75&0.89&212\\
LaborForce&$1.02\pm0.03$&0.97&1.00&205\\
AgriculturalValueAdded&$1.01\pm0.05$&0.88&0.95&207\\
HouseholdConsumption&$1.03\pm0.11$&0.64&0.82&206\\
TotalConsumption&$1.01\pm0.11$&0.62&0.79&208\\
\\
CarbonEmission&$1.30\pm0.17$&0.60&0.95&153\\
\\
CellularPhones&$1.17\pm0.09$&0.78&1.02&211\\
InternetUsers&$1.16\pm0.13$&0.61&0.88&208\\
\\
RoadLength&$1.00\pm0.06$&0.82&0.90&210\\
\hline
\end{tabular}
\end{table}

\end{center}

One of the obvious facts is all properties (except boundary length)
have super linear or linear allometric scaling relationships with
respect to population. We know that the geographic, communication
related and energy related properties have super linear relation as
well as the economic, trade related and transportation properties
have approximately linear relations with respect to population. Many
economic-related properties such as GDP, government debt are
filtered out because their $r^2$s are smaller than $0.6$. However,
we know that the economic-related properties such as GDP, number of
new patents always have super-linear relation with population in
cities\citep{Bettencourtwest2007}. To make the differences between
cities and countries clearer, we list the comparisons of several
properties with respect to population between cities and countries
in table \ref{tab.comparison}.

\begin{center}
\small
\begin{longtable}{p{4cm}p{3cm}p{2cm}p{1cm}p{2.5cm}}
\caption{Comparison of Countries and Cities in Allometric Scaling
Relations with respect to Population. ($\alpha$,observations and
$r^2$ are parameters of countries. $\alpha$ in cities are from
\citep{Bettencourtwest2007} and \citep{Lee1989}.)} \label{tab.comparison}\\

\hline Property &$\alpha$ in cities&$\alpha$&$r^2$&Observations
\\
\hline
\endfirsthead

\hline \multicolumn{4}{r}{{Continued on next page}} \\ \hline
\endfoot
\hline
\endlastfoot
Area&0.33-0.91&$1.30\pm0.11$&0.71&212\\
GDP&1.13-1.26&$1.04\pm0.12$&0.59&212\\
Electricity Consumption&1.07&$1.45\pm0.19$&0.54&202\\
HIV AIDS Population&1.23(New aids cases)&$1.88\pm0.37$&0.40&163\\
Household Consumption&1.00-1.05(Household electricity/water consumption)&$1.03\pm0.11$&0.64&206\\
Oil Consumption&0.79(Gasoline sales)&$1.05\pm0.13$&0.56&201\\
\hline
\end{longtable}
\end{center}

We can see that the exponents of countries are very different from
cities. Especially, the exponents of GDP (which is not appear in
table \ref{tab.population} because its $r^2$ is smaller than 0.6) is
around 1. That means the economic activities per capita doesn't
increase with population. The electricity consumption and oil
consumption have larger exponents than cities which means the energy
consumption per capita increases with population in countries.

Another significant phenomenon in table \ref{tab.comparison} is the
variances(the width of credit interval) of $\alpha$ are very large,
as well as the coefficients of determination are small in countries.
(The variances of exponents in cities are much more smaller than
countries according to \citep{Bettencourtwest2007}). That means
there are large uncertainties on exponent estimations in countries.
Therefore, population may be not an appropriate proxy of size of
countries as cities. We will further discuss this problem in
sub-sections \ref{sec.GDP} and section \ref{sec.discussion}.

\subsection{Allometric Scaling with respect to GDP}
\label{sec.GDP}

We study the allometric scaling relations with respect to economic
sizes of the countries by the same method, and the results are shown
in table \ref{tab.gdp}.
\begin{center}
\small
\begin{longtable}{p{5cm}p{2cm}p{1cm}p{1cm}p{2.5cm}}
\caption{Allometric Scaling Relations with respect to GDP(Exponent
$\alpha$ is estimated by MA method and $\alpha'$ is
 estimated by OLS method). Only the allometries with $r^2$ larger than 0.6 are shown.
 GDP data in 1998 is used to show the relations of Radio Stations and Television Stations with respect to GDP because we have only the data of these two properties in 1998 (See appendix)}
 \label{tab.gdp} \\
      \hline
Property &$\alpha$&$r^2$&$\alpha'$&Observations
\\ \hline
\endfirsthead

\hline \multicolumn{4}{r}{{Continued on next page}} \\ \hline
\endfoot

\hline
\endlastfoot

GDPAtParity&$1.02\pm0.03$&0.96&0.99&230\\
GovernmentDebt&$1.10\pm0.09$&0.85&1.00&120\\
GovernmentExpenditures&$1.02\pm0.04$&0.93&0.98&224\\
GovernmentReceipts&$1.04\pm0.04$&0.93&1.00&225\\
GovernmentSurplus&$1.06\pm0.07$&0.78&0.93&217\\
LaborForce&$1.02\pm0.09$&0.70&0.85&223\\
NationalIncome&$0.99\pm0.01$&1.00&0.99&208\\
AgriculturalValueAdded&$0.96\pm0.08$&0.73&0.83&207\\
ConstructionValueAdded&$0.99\pm0.03$&0.96&0.97&208\\
FixedInvestment&$0.98\pm0.02$&0.97&0.97&207\\
GovernmentConsumption&$1.00\pm0.04$&0.91&0.96&207\\
HouseholdConsumption&$0.98\pm0.02$&0.98&0.97&206\\
IndustrialValueAdded&$1.18\pm0.03$&0.96&1.15&208\\
InventoryChange&$1.26\pm0.13$&0.67&0.99&193\\
ManufacturingValueAdded&$1.22\pm0.04$&0.95&1.19&208\\
MiscellaneousValueAdded&$1.05\pm0.03$&0.96&1.03&208\\
TotalConsumption&$0.97\pm0.02$&0.98&0.96&208\\
TradeValueAdded&$1.00\pm0.02$&0.97&0.98&207\\
TransportationValueAdded&$0.99\pm0.03$&0.96&0.97&207\\
\\
ExportValue&$1.09\pm0.05$&0.90&1.03&208\\
ImportValue&$0.95\pm0.04$&0.92&0.92&209\\
CurrentAccountBalance&$0.97\pm0.07$&0.84&0.89&163\\
ForeignExchangeReserves&$1.00\pm0.08$&0.80&0.89&155\\
ExternalDebt&$1.17\pm0.07$&0.86&1.07&201\\
\\
ElectricityConsumption&$1.25\pm0.09$&0.80&1.09&215\\
ElectricityProduction&$1.16\pm0.07$&0.85&1.05&213\\
OilConsumption&$0.95\pm0.05$&0.89&0.90&211\\
OilImports&$0.92\pm0.07$&0.77&0.81&200\\
CarbonEmission&$1.10\pm0.06$&0.89&1.03&153\\
\\
PhoneLines&$0.98\pm0.05$&0.88&0.92&229\\
CellularPhones&$1.14\pm0.07$&0.83&1.03&223\\
RadioStations*&$0.66\pm0.06$&0.69&0.58&205\\
TelevisionStations*&$0.75\pm0.08$&0.65&0.64&197\\
InternetUsers&$1.06\pm0.06$&0.85&0.97&221\\
\\
RoadLength&$0.95\pm0.09$&0.69&0.80&223\\
\hline
\end{longtable}
\end{center}

Interestingly, most economic related properties have exponents
around 1, they increase proportionally with GDP. However, some
properties such as government debt, industrial value added,
inventory change, manufacturing added values have super-linear
relations with respect to GDP because their exponents are larger
than 1 significantly.

In trade-related properties, export value and external debt have
super-linear relations, while import value has sub-linear relation
with respect to GDP. If we treat the economy of a whole country as a
money flow system, then GDP is the body size of this system, and
import value dedicating money out-flow is just the metabolism of the
whole system. Hence the allometric scaling relationship between
import value and GDP is the correspondence of Kleiber law in money
flow\citep{kleiber1932}. However, the fact that the exponent of this
allometry is 0.95 needs more careful explanations.

For the energy-related properties, the electricity consumption,
production and the carbon emission which are important measures of
the metabolism of countries grow faster than GDP. However, the oil
consumption and import all have sub-linear allometric scaling
relations with respect to GDP. That means the economy depends more
on electricity power rather than gas or oil consumptions as GDP
increases.

Similarly, in the category of communication, radio stations,
television stations, phone lines have sub-linear relation with
respect to GDP, nevertheless, cellular phones, internet hosts and
users have super linear or linear relations with respect to GDP
which also means the variables with super linear relations grows
faster than those with sub linear relations.

\subsection{Population and GDP}
\label{sec.PopulationGDP}

Among the allometries we have discussed in previous sections, the
most important one is the allometry between population and GDP
because this relation links not only natural properties and economic
characteristics of countries but also demographic size and economic
size of countries. However, this allometry does not appear in table
\ref{tab.population} because the value of $r^2$ between these two
variables is very small. Is it possible to improve the fitting
goodness? We will discuss this problem in this sub-section.

\begin{figure}
\centerline {\includegraphics{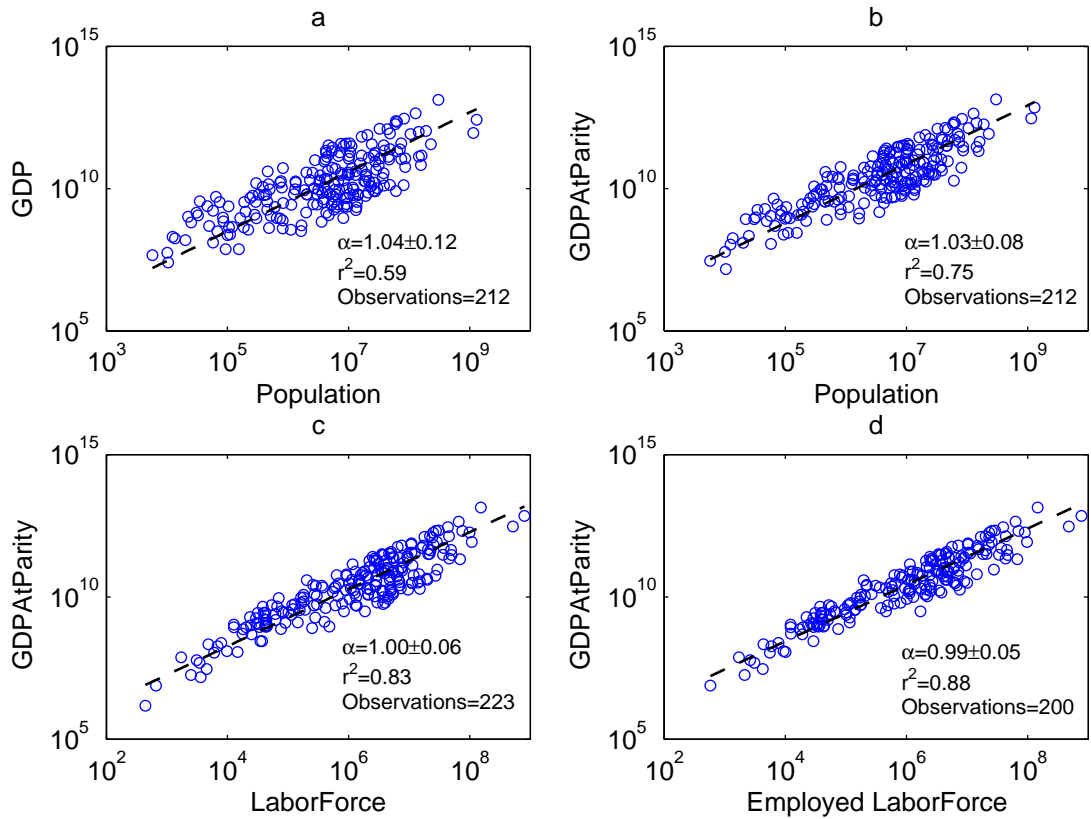}} \vskip3mm \caption{
Improvements of Allometric Scaling Relations between Population and
GDP}\label{fig.populationgdp}
\end{figure}

At first, we can use GDP at parity instead of GDP because GDP at
parity measures the real purchasing power of one economy. More
improvement can be made by using labor force property instead of the
population property because the labor force fraction of population
can really contribute to the macro-economy. And further improvement
can be made by calculating the employed labor force (labor force
$\times$ (1-unemployment fraction)) rather than labor force. These
results are shown in figure \ref{fig.populationgdp}.

As shown in figure \ref{fig.populationgdp} from (a) to (d), data
become more concentrated with larger $r^2$ values when population
and GDP are replaced by more appropriate properties employed labor
force and GDP at parity. Meanwhile, the exponent also decreases and
converges to 1 when suitable properties are used. We can conclude
that GDP increases proportional with population. This is very
different from the super-linear relations in cities
\citep{Bettencourtwest2007,chen2003}.

\subsection{Allometric scaling of intensive variables}
\label{sec.intensive}

Although we have investigated allometric scaling of various
variables with different types of body size of countries, all
variables are just extensive ones. According to statistical physics,
the intensive variables can reflect the state of systems in average.
Here, we also show the allometries of intensive variables.
\begin{table}
\centering
 \caption{Allometric Scaling Relations between Intensive Variables and GDPPerCapita}
 \label{tab.intensive}
\begin{tabular}{ccccccc}
\hline
Property &$\alpha$  & $r^2$ & $\alpha'$&Observations\\
\hline

LiteracyFraction&$0.11\pm0.02$&0.40&0.11&204\\
BirthRateFraction&$-0.24\pm0.03$&0.58&-0.23&208\\
LifeExpectancy&$0.08\pm0.01$&0.48&0.08&207\\
MedianAge&$0.14\pm0.02$&0.63&0.14&208\\
TotalFertilityRate&$-0.21\pm0.03$&0.49&-0.20&208\\
InfantMortalityFraction&$-0.58\pm0.05$&0.71&-0.53&206\\

\hline
\end{tabular}
\end{table}

\begin{figure}
\centerline {\includegraphics{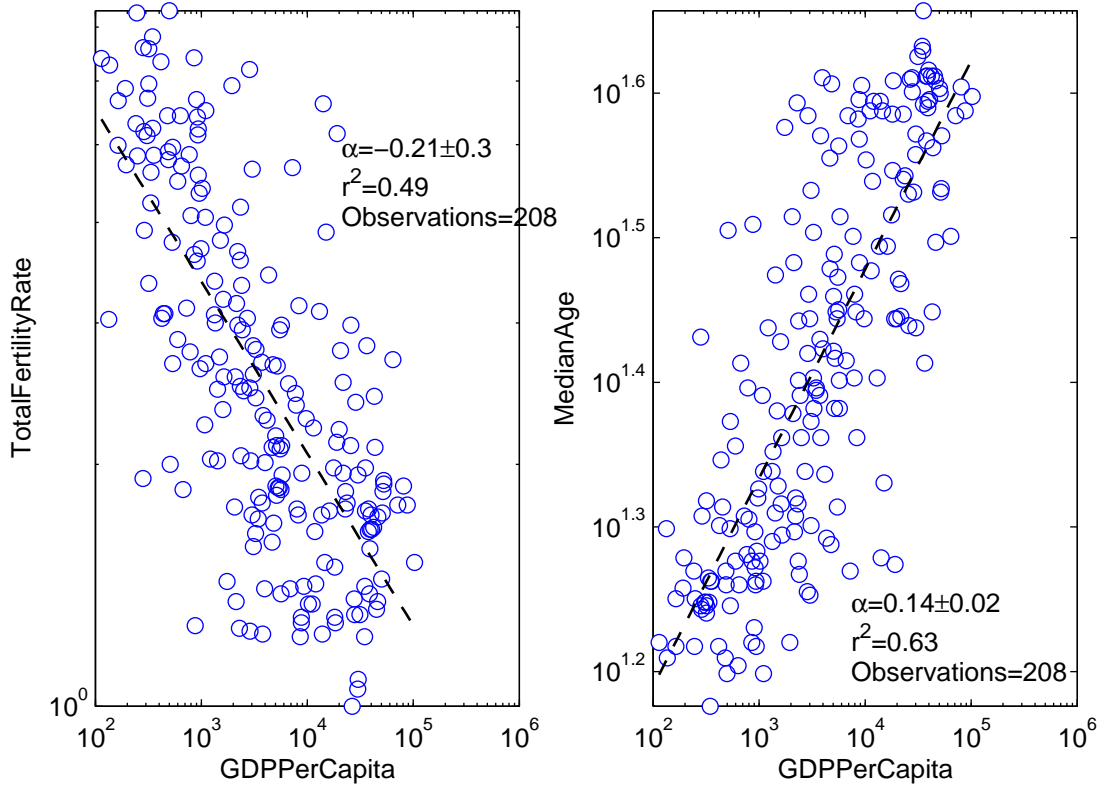}} \vskip3mm \caption{
Allometric Scaling Relations of TotalFertilityRate, MedianAge v.s.
GDPPerCapita}\label{fig.intensive}
\end{figure}

We investigate the allometric scaling of intensive variables with
respective to GDP per capita. Some intensive properties have scaling
relation with respect to GDP per capita(table \ref{tab.intensive}).
Two of variables are selected to plot in figure \ref{fig.intensive}.

The relations between birth rate fraction, median age and GDP per
capita deserve more attention. In \citep{Moses2006}, the authors
predicted that the power law relation between fertility and energy
consumption per capita for U.S. in different years has the exponent
-1/3 according to the Kleiber's law. Energy consumption per capita
was regarded as the proxy of metabolism of humans in their
work\citep{Moses2006}. In fact, we didn't find the relevant fact by
using energy consumption per capita (we've tested Oil consumption,
electrical consumption and carbon emission per capita). However, GDP
per capita as the important economic index may have a deep
connection with metabolism of individuals.

\section{Discussion}
\label{sec.discussion}
\subsection{Comparisons between Regression Methods}
In this paper, the main conclusions are drawn according to the major
axis regression method. However, we also listed the estimated
allometric scaling power law exponents in tables \ref{tab.area},
\ref{tab.population}, \ref{tab.gdp} and \ref{tab.intensive}. Because
MA method tries to minimize the total distances from data to the
regression line in both directions but not only the $y$ direction,
it considers the uncertainties both from independent variable and
dependent variable\citep{warton2006}. The difference between
$\alpha$ and $\alpha'$ decrease with the correlation degree between
two variables which is measured by $r^2$. Sometimes the super linear
or linear allometric scaling relations may be underestimated by OLS
method as sub linear relations (see table \ref{tab.population}) when
$r^2$ is large.

Another advantage of MA method is it always gives compatible
exponent estimations when independent variable and dependent
variable are interchangeable. For example, the power law relation
between area and population in table \ref{tab.area} is $Y\sim
X^{0.77}$, where $Y$ stands for population and $X$ stands for area.
In table\ref{tab.population}, we have another power law relation,
$X\sim Y^{1.30}$. These two allometric scaling relations are
compatible because the product of exponents $0.77\times
1.30=1.001\approx 1$. However, we know that $\alpha'$ is $0.67$ in
table \ref{tab.area}, and $1.05$ in \ref{tab.population}. But, the
product of these two exponents $0.67*1.05=0.7035\approx 0.71=r^2$.
Therefore, we can not get compatible exponents by OLS method.

As we know, the OLS method assumes that the dependent variable,
i.e., $log(M)$ is measured without errors to give a reasonable
estimation of $\alpha$. However, the purpose of this paper is to
find possible allometric scaling relations between macro-variables
of countries but not to predict some properties according to a given
allometric scaling relation, so MA method is more suitable than OLS
method. Second, MA method is symmetric and regardless of which
variable is independent variable and which is dependent one. In this
paper, it is necessary to compare one pair of variables with
different orders. So MA method is preferable\citep{warton2006}.
However, we should also point out that MA method can not be abused
anywhere because it may have its own limitation that is the
variances of dependent and independent variables should be
equal\citep{smith2009}. In this paper, we assume this condition is
satisfied, that is most variables have similar measurement errors.
Therefore, MA method can be applied to give more appropriate
estimations.

\subsection{Similarities and Differences between Cities and Countries}
Allometric scaling is a very ubiquitous law in complex systems. Lots
of macro-variables exhibit power law relations both in cities and
countries. It implies that some common mechanisms govern the
development of these complex systems. There are several possible
explanations for universal power law relations in the literatures.
First, self-affine fractals may explain the allometric scaling
between two variables. As pointed by \cite{Chen2009,Chen2008},
cities are self-affine fractals, therefore, the allometric scaling
exponent between two variables is the ratio of two fractal
dimensions. This conclusion may also suitable for countries.

Another possible explanation is that self-similar transportation
networks may exist widely in organisms, cities and
countries\citep{banavar1999, Isalgue2007, Kuhnert2006}. Supply
networks such as traffic networks and electricity grids resembling
the vascular networks in organisms determine the power law relations
between size of cities or countries and energy
consumption\citep{Kuhnert2006,zhang2009}. However, we find it is
week to explain other ubiquitous allometric scaling relations, such
as population and GDP, where the transportation networks are hard to
be found.

Finally, we assert another kind of possible mechanism can explain a
large varieties power law relations both in cities and countries.
The ubiquitous power law relations imply different cities or
countries may have similar structures which can be described by
similar distributions. For example, there are common wealth
distribution patterns in different countries\citep{dragulescu2001}.
The similar distribution curves may determine the power law relation
between population and GDP. Some studies have found the relation
between power law distributions and relations in
language\citep{bernhardsson2009} and family
names\citep{miyazima2000}.

However, there are many differences between cities and countries.
First, the allometric scaling relations in countries are less
significant than cities because countries as units of research are
more different and heterogenous. For example, some developing
countries with large population are less urbanized than developed
countries, therefore, their economic structures, cultural
conventions and living styles are very different. However, in
studies of cities, researchers always select cities with similar
economic or geographic
conditions\citep{batty1994,Bettencourtwest2007}. That is the reason
why the allometries in countries are more uncertain and less
significant than cities.

Second, the exponent of the same allometric scaling relations are
different between countries and cities. For example, in table
\ref{tab.comparison}, the exponent of population and GDP is $1.04$
for countries. However, it is about $1.20$ for cities. One of the
possible explanations is cities are more free for the flows than
countries. As we all know, people, money and materials can flow
freely and quickly between different cities in one country.
Consequently even the boundaries of cities can change
dynamically\citep{Chen2008,batty1994}. However, people, money and
goods flows are much weaker between countries. And the boundaries of
countries are almost unchanged. Hence countries are rigid to self
organize. This may explain the differences of exponents between
cities and countries.

\section{Concluding Remarks}

In this paper, we used MA method to find the allometric scaling
relationships among numerical macro-variables in different
countries. By systematic studies, we found some interesting facts.

In general, the results reveal that economic, trade, energy
consumption and communication related properties have significant
allometric scalings with respect to GDP of countries (see table
\ref{tab.gdp}). And the geographic, demographic, and natural
resource related properties have significant scaling relations with
respect to area or population of countries (see table \ref{tab.area}
and \ref{tab.population}).

Many differences between countries and cities are revealed by the
studies of allometric scaling relations with respect to population.
One reason is countries are more heterogenous than cities. The other
possible reason is the mobility between cities is much larger than
countries.

We also studied the allometric scaling between intensive variables
and GDP per capita. Some regularities and connections with metabolic
theory are pointed out.

\paragraph{Acknowledgement}
We acknowledge Wolfram Research Inc. to provide the software
Mathematica 7 and the country data. The author thanks three
anonymous reviewers to give constructive suggestions. This research
was supported by National Natural Science Foundation of China under
Grants of No. 70771012

\appendix
\section{Demonstration of Properties}
\label{sec.appendix}
 Appendix Table \ref{tab.appendix} lists properties of
countries. The first column is the name of the property in
Mathematica, the second column is the explanation of the property.
The third column is its category of the property (Totally 11
categories). And the fourth column denoting whether the property is
intensive (I) or extensive (E) one. The last column is the time (in
which year) that the property refers to. The stars behind the
numbers indicate that most but not all data of this property are
from one single year for all countries. All these properties can be
obtained from Mathematica except property of "carbon emission".
Actually, there are totally $225$ different properties in
"CountryData" command of Mathematica. However, most of properties,
such as the flag, or the capital city of a country, are not
numerical. So we selected only numerical properties to study.
Furthermore, because some numerical properties are so similar that
they can give identical allometric scaling relations (such as
military age male population and military age female population,
etc.), we only selected some of the representative properties to
study. Finally, 92 properties which appeared in the main text are
shown.

\begin{center}
\small
\begin{longtable}{|p{3cm}|p{4.5cm}|p{3cm}|p{0.5cm}|p{0.5cm}|}
\caption{List of Country Properties} \label{tab.appendix} \\
      \hline
\multicolumn{1}{|c|}{\textbf{Property}} &
\multicolumn{1}{c|}{\textbf{Explanation}} &
\multicolumn{1}{c|}{\textbf{Category}}&
\multicolumn{1}{c|}{\textbf{E/I}}&
\multicolumn{1}{c|}{\textbf{Year}}
\\ \hline
\endfirsthead

\hline \multicolumn{5}{|r|}{{Continued on next page}} \\ \hline
\endfoot

\hline
\endlastfoot

Area & Total country area in square kilometers & Geographical  & E & -\\
Boundary Length & Total large-scale boundary length in kilometers & Geographical & E & -\\
ArableLand Area & Arable land area in square kilometers & Natural resources and features & E & 2005\\
IrrigatedLand Area & Irrigated land area in square kilometers & Natural resources and features & E & 2003*\\
Natural Resources & Number of primary natural resources & Natural resources and features & E &-\\
Population & Estimated population & Demographic & E & 2006\\
HIV AIDS Population & Total population infected with HIV & Public health & E & 2003*\\
Military Fit Population & Total population considered fit for military service & Military-related & E& 2005\\
GDP & GDP in U.S. dollars at official exchange rate & Economic-related & E & 2006\\
GDP At Parity & GDP in U.S. dollars at purchasing power parity & Economic-related & E& 2007*\\
Government Debt & Outstanding government debt in U.S. dollars & Economic-related & E& -\\
Government Expenditures & Annual government expenditures in U.S. dollars & Economic-related & E& 2005* \\
Government Receipts & Annual government receipts in U.S. dollars & Economic-related & E& 2005* \\
Government Surplus & Annual government surplus in U.S. dollars & Economic-related & E& 2005*\\
Labor Force & Size of adult labor force, whether employed or not & Economic-related & E& 2007*\\
National Income & National income in U.S. dollars at official exchange rate & Economic-related & E&2006\\
Agricultural Value Added & Value added by agricultural activities & Economic-related & E&2006\\
Construction Value Added & Value added by construction and real estate activities & Economic-related & E & 2006\\
Fixed Investment & Investment in fixed capital & Economic-related & E & 2006\\
Government Consumption & Annual government consumption & Economic-related & E & 2006\\
Household Consumption & Annual household consumption & Economic-related & E & 2006\\
Industrial Value Added & Value added by all industrial activities & Economic-related & E & 2006\\
Inventory Change & Annual change in the value of inventories & Economic-related & E & 2006\\
Manufacturing Value Added & Value added by manufacturing industries & Economic-related & E&2006\\
Miscellaneous Value Added & Value added by miscellaneous service and other activities & Economic-related & E & 2006\\
Total Consumption & Total consumption expenditure & Economic-related & E & 2006\\
Trade Value Added & Value added by wholesale and retail trade & Economic-related & E & 2006\\
Transportation Value Added & Value added by transportation and communications & Economic-related & E & 2006\\
ExportValue & Total estimated value of annual exports in U.S. dollars & Trade-related & E &2006\\
ImportValue & Total estimated value of annual imports in U.S. dollars & Trade-related & E &2006\\
Current Account Balance & Current account trade balance in U.S. dollars & Trade-related & E& 2007*\\
Foreign Exchange Reserves & Government reserves of currency and gold in U.S. dollars & Trade-related & E& 2007*\\
External Debt & Total government foreign currency debt, in U.S. dollars & Trade-related & E& 2007*\\
Electricity Consumption & Annual electricity consumption in kilowatt hours & Energy-related & E& 2005*\\
Electricity Production & Annual electricity output in kilowatt hours & Energy-related & E& 2005*\\
Oil Consumption & Oil consumption in barrels per day & Energy-related & E& 2005*\\
Carbon Emission & Total anthropogenic carbon dioxide emissions(Gg) & Energy-related & E & 2005\\
Phone Lines & Number of telephone land lines in use & Communications-related & E& 2006*\\
Cellular Phones & Number of cellular phones in use & Communications-related & E& 2006*\\
Radio Stations & Total number of public radio stations & Communications-related & E& 1998*\\
Television Stations & Number of broadcast television stations & Communications-related & E&1997*\\
Internet Users & Estimated number of internet users & Communications-related & E& 2006*\\
Airports & Total number of airports & Transportation-related & E&2007\\
Road Length & Total length of all roads in kilometers & Transportation-related & E& 2004* \\
Literacy Fraction & Fraction of adult population able to read and write & Culture-related & I& 2003*\\
Population Growth & Annual fractional growth in estimated population & Demographic & I& 2005*\\
Birth Rate Fraction & Average number of births per person per year & Demographic & I& 2008*\\
Life Expectancy & Average life expectancy in years & Demographic & I& 2008* \\
Median Age & Median age of population in years & Demographic & I& 2006\\
Total Fertility Rate & Lifetime total average number of births per woman & Demographic & I& 2008* \\
Infant Mortality Fraction & Fraction of births with infant mortality & Public health & I& 2008* \\
GDP PerCapita & GDP normalized by population & Economic-related & I& 2006\\
Unemployment Fraction & Fraction of labor force unemployed & Economic-related & I& 2007*\\

\end{longtable}
\end{center}


{\textbf{References}} 


\bibliographystyle{elsarticle-num}
\bibliography{ecology}







\end{document}